\documentclass[twocolumn]{aastex63} 
\usepackage{amsmath,  amssymb, url, graphicx}
\usepackage{balance}
\renewcommand{\vec}[1]{\pmb{#1}}

\newcommand{\beq}{\begin{equation}}
\newcommand{\eeq}{\end{equation}}

\newcommand{\T}{{\cal T}}

\newcommand{\M}{{\cal M}}

\shorttitle{Radio Emission and Electric Gaps in Pulsar Magnetospheres}

\begin{document}

\title{Radio Emission and Electric Gaps in Pulsar Magnetospheres}

\correspondingauthor{Ashley Bransgrove}
\email{abransgrove@princeton.edu}

\author{Ashley Bransgrove} 
\affil{Physics Department and Columbia Astrophysics Laboratory, Columbia University, 538 West 120th Street, New York, NY 10027}
\affil{Princeton Center for Theoretical Science and Department of Astrophysical Sciences, Princeton University, Princeton, NJ 08544, USA}

\author{Andrei M. Beloborodov}
\affil{Physics Department and Columbia Astrophysics Laboratory, Columbia University, 538 West 120th Street, New York, NY 10027}
\affiliation{Max Planck Institute for Astrophysics, Karl-Schwarzschild-Str. 1, D-85741, Garching, Germany}

\author{Yuri Levin}
\affil{Physics Department and Columbia Astrophysics Laboratory, Columbia University, 538 West 120th Street, New York, NY 10027}
\affil{Center for Computational Astrophysics, Flatiron Institute, 162 5th Avenue, 6th floor, New York, NY 10010
}
\affil{Department of Physics and Astronomy, Monash University, Clayton, VIC 3800, Australia}



\begin{abstract}
The origin of pulsar radio emission is one of the old puzzles in theoretical astrophysics. In this Letter we present a global kinetic plasma simulation which shows from first-principles how and where radio emission can be produced in pulsar magnetospheres. We observe the self-consistent formation of electric gaps which periodically ignite electron-positron discharge. The gaps form above the polar-cap, and in the bulk return-current. Discharge of the gaps excites electromagnetic modes which share several features with the radio emission of real pulsars. We also observe the excitation of plasma waves and charge bunches by beam instabilities in the outer magnetosphere. Our numerical experiment demonstrates that global kinetic models can provide deep insight into the emission physics of pulsars, and may help interpret their multi-wavelength observations.  
\end{abstract}


\section{Introduction}
Pulsars are rotating magnetized neutron stars which produce powerful beams of coherent radio emission. Despite an abundance of observational data, the mechanism generating radio waves in pulsar magnetospheres has remained elusive for more than fifty years. Pulsar magnetospheres are filled with highly magnetized collisionless electron-positron ($e^{\pm}$) plasma which is produced in electric gaps --- regions with voltage along magnetic field lines. Electric discharge in the gaps may be responsible for pulsar radio emission, and the only reliable way to solve this nonlinear 
problem is with a self-consistent numerical simulation. 

In the last decade there has been a significant computational effort to model the magnetosphere structure and multi-wavelength emission from first-principles using the particle-in-cell (PIC) technique  \citep{chen_electrodynamics_2014,philippov_ab-initio_2015,philippov_ab-initio_2018}. Global simulations unanimously display powerful gamma-ray emission from the outer magnetosphere \citep{ gruzinov_2013, chen_electrodynamics_2014, philippov_ab-initio_2018}, but radio waves were not observed. Recently \cite{philippov_origin_2020} showed that radio waves can be generated directly by the  
$e^{\pm}$ discharge above the polar-cap. The authors used local PIC simulations of the gap in Cartesian geometry --- it required high voltage and resolution which were not achieved in global simulations. 

In this Letter we present a global kinetic plasma simulation of an axisymmetric
magnetosphere. Our numerical experiment achieves voltage and spatial resolution 
sufficient
to reveal 
new features,
including locations of electric gaps where $e^{\pm}$ discharge 
ignites and produces
coherent radio waves. 
We observe
beam 
instabilities in the 
magnetosphere
for the first time.

\begin{figure*}[t]
\centering
\includegraphics[width=\textwidth]{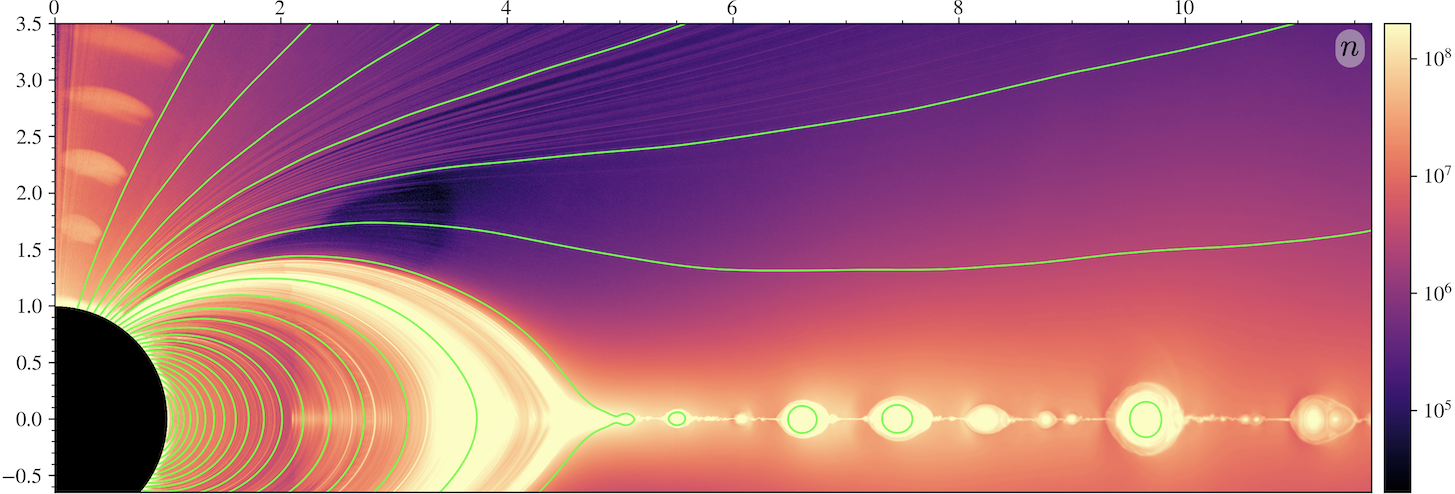} 
\caption{
Poloidal cross section of the magnetosphere
in the quasi-steady state.  
Green curves show poloidal magnetic field lines,
and background color shows plasma density.
The 
clouds of plasma above the polar cap form in $e^{\pm}$ discharges. 
The absence of plasma (dark zone) above the separatrix shows the return-current gaps in the unscreened state; it fills with plasma every $\sim 2 R_\text{LC}/c$ during discharge episodes. Magnetic reconnection forms a dynamic plasmoid chain in the equatorial current sheet.
}
\label{global}
\end{figure*}

\section{Method} 
The kinetic plasma simulation is performed with the relativistic PIC code {\tt{Pigeon}} \citep{ hu_relativistic_2021,hu_axisymmetric_2021}, a descendant of {\tt{Aperture}} \citep{chen_electrodynamics_2014},
which solves Maxwell's equations 
with the equations of motion for charged particles and $e^\pm$ creation.
We include a general-relativistic correction to Faraday's law which describes the rotational frame-dragging effect near the star \citep{philippov_ab-initio_2018}.
We set the compactness of the star $r_s/r_\star=0.5$, where $r_s=2GM/c^2$ is the Schwarzschild radius and $r_\star$ is the radius of the star. 
Hereafter lengths are given in units of $r_\star$, and times in units of $r_\star/c$. The $e^{\pm}$ have mass $m$ and charge $\pm e$, while ions (protons) have charge $e$ and mass $m_i=10m$. Electromagnetic fields are given in units of $mc^2 e^{-1}r_\star^{-1}$, number density in units of $r_\star^{-3}$, and charge density in units of $e r_\star^{-3}$.

The scales of the problem are set by the angular velocity $\Omega$ and the magnetic dipole moment $\mu$ of the pulsar. Rotation of the magnetized star induces a voltage $\Phi_0\approx\mu \Omega^2 / c^2$ 
capable of accelerating $e^\pm$ to 
Lorentz factor $\gamma_0 = e\Phi_0 / m c^2$. 
For real pulsars $\gamma_0\sim 10^{10}$. 
We scale it down 
to 
$\gamma_0=2\times 10^4$, which allows us to resolve the plasma skin depth. Our value of $\gamma_0$ is twice as large as 
previous work \citep{hu_axisymmetric_2021},
and our resolution is twice as high. We set $\Omega=1/6$, which implies the light-cylinder radius $R_\text{LC}=c/\Omega = 6$. The minimum charge density required to support the co-rotating magnetosphere
$\rho_\text{GJ}=-\vec{\Omega}\cdot \vec{B}/(2\pi c)$ \citep{goldreich_pulsar_1969}
defines a
characteristic number density $n_\text{GJ}=|\rho_\text{GJ}|/e$ and 
a
plasma frequency 
$\omega_p=(4\pi n_\text{GJ} e^2/m)^{1/2}$. 
Similar to real pulsars, the simulation preserves
the hierarchy of scales $ \Omega \ll \omega_p \ll \omega_B$,
where $\omega_B=eB/m c$ is the cyclotron frequency.

The gamma-ray emission, propagation, and $e^{\pm}$ production is modelled using the Monte-Carlo method \citep{chen_electrodynamics_2014}. The $e^{\pm}$ emit gamma-rays of energy $\epsilon_\gamma=10mc^2$ 
when they reach the threshold energy $\gamma_\text{thr}mc^2$. The threshold depends on the 
curvature of the field lines \citep{chen_electrodynamics_2014}, 
and has typical value
$\gamma_\text{thr}\sim 100$.
The gamma-rays propagate with mean free path 
$\bar{\ell}$
before converting to secondary $e^{\pm}$ with $\gamma_s mc^2 = \epsilon_\gamma/2$. For $r<2$, we set 
$\bar{\ell}$
to simulate conversion off the strong magnetic field near the stellar surface ($\gamma-B$ channel). For $r>2$,  we set 
$\bar{\ell}$
to simulate collisions with soft target photons in the outer magnetosphere ($\gamma-\gamma$ channel). The energy scales in our 
numerical experiment
satisfy the correct hierarchy $1 \ll \gamma_s \ll \gamma_\text{thr} \ll \gamma_0$. 

The computational domain covers $1 \leq r \leq 30$ and $0 \leq \theta \leq \pi$. The grid is uniform in log~$r$ and $\theta$, and has resolution $N_r\times N_\theta = 8192\times 8192$ cells. The 
plasma scale  
$2\pi c/\omega_p$
near the 
stellar
surface is resolved by $25$ grid cells.
At the surface, electromagnetic
fields satisfy a rotating conductor boundary condition, and 
we maintain a gravitationally bound electron-ion atmosphere with multiplicity $\mathcal{M}_{atm}=n_{atm}/n_\text{GJ}=10$.
At the outer boundary 
fields are damped, and particles are absorbed. 

\begin{figure*}[t!]
\centering 
\includegraphics[width=\textwidth]{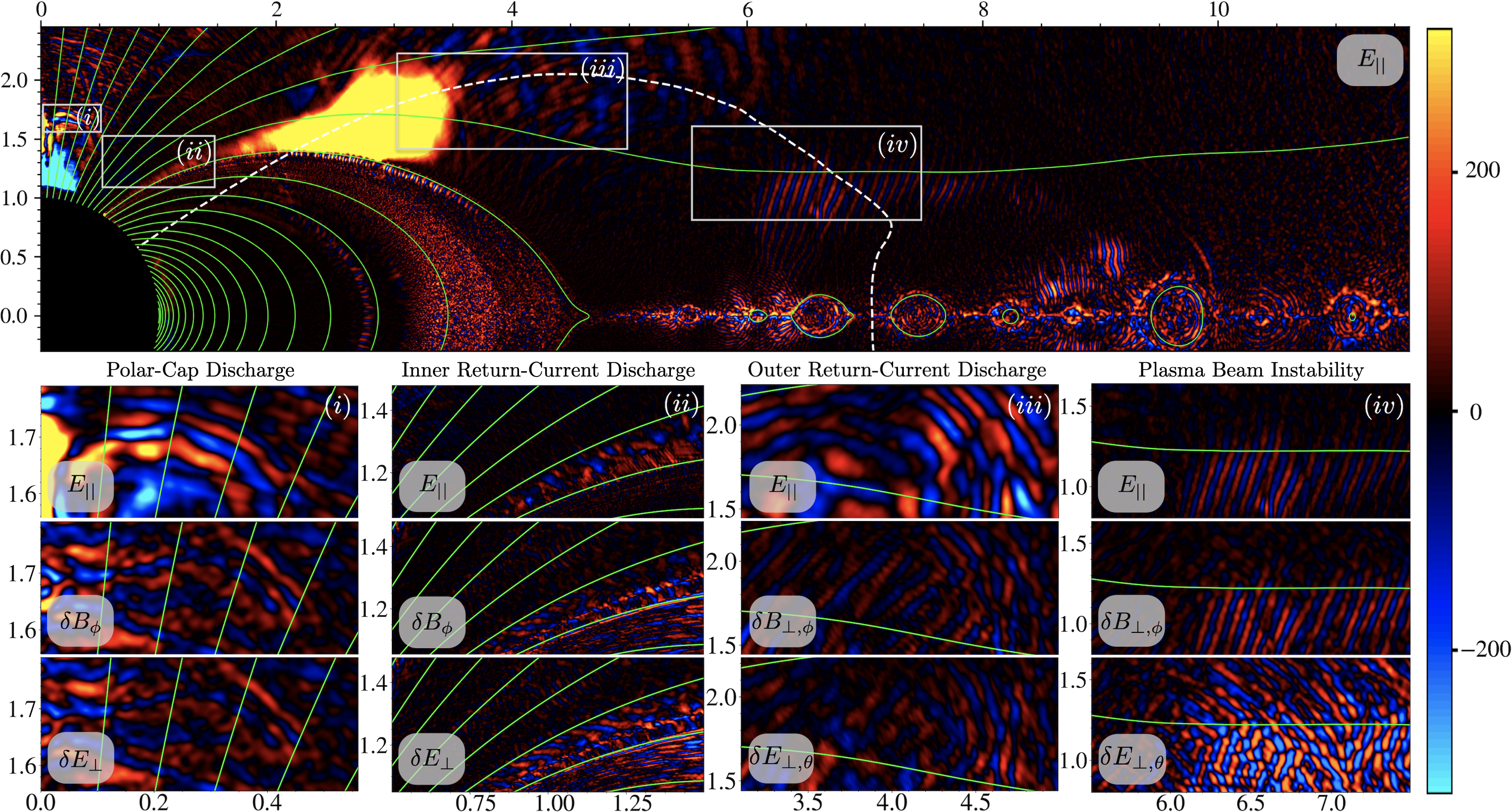} 
\caption{
Excitation 
of waves in the pulsar magnetosphere. Color
bar shows
the
components of the electromagnetic field multiplied by $(r/r_\star)^2$.
Top : $E_{||}$ in the quasi-steady state. The dashed white curve shows the null-surface $\rho_\text{GJ}=0$ (the null-surface breathes outside $r_\text{LC}$ due to plasmoid motion). Insets (i)-(iii): Electromagnetic modes excited during $e^{\pm}$ discharges at the polar-cap, inner return-current, and outer return-current gaps. Inset (iv): Waves excited by a 
beam plasma instability. For insets (i) and (ii), $\delta E_\perp$ is calculated by projecting $\delta \vec{E}_\perp$ onto a unit vector in the $\vec{\hat{\phi}}\times \vec{B}$ direction.}
\label{waves}
\end{figure*}

\section{Magnetosphere Structure} 
The simulation begins with 
a non-rotating star with a vacuum
dipole 
magnetosphere. 
We smoothly increase the angular velocity from zero 
to $\Omega$ during the first $5$~$r_\star/c$, and observe that charges are lifted from the atmosphere 
and accelerated, triggering gamma-ray emission and $e^{\pm}$ production which fills the magnetosphere with plasma. The magnetosphere reaches a quasi-steady state after the first revolution
of the star (Fig.~\ref{global}). 
Its 
global structure is similar to previous axisymmetric force-free and PIC models \citep{Contopoulos_1999,Gruzinov_2005,parfrey_introducing_2012, chen_electrodynamics_2014}.
The higher voltage in our simulation leads to more efficient filling of the open field lines with plasma. A new feature is the 
$e^\pm$ discharge in the outer magnetosphere outside the Y-shaped current sheet. This feature was not seen
in previous PIC simulations with lower
voltages.
We also observe $e^\pm$ discharge near the polar caps, an expected result  of the relativistic frame-dragging effect \citep{muslimov_tsygan}. The strongest dissipation and gamma-ray emission 
occur
in the equatorial current sheet.

\section{Gaps and Electric Discharge} 
The gaps 
(regions with non-zero $E_{||}=\vec{E}\cdot\vec{B}/|\vec{B}|$) form because of
a mismatch between the required parallel current $j_B= (c/4\pi)(\nabla\times\vec{B})\cdot\vec{B}/|\vec{B}|$ and the 
maximum 
current 
$j_\text{max} = c \rho_\text{GJ}$ that could
be supplied by the outward flow of a charge-separated plasma with the local charge density $\rho=\nabla\cdot\boldsymbol{E}/4\pi$.
If the charged plasma under-supplies the required current ($\alpha\equiv j_B / j_\text{max}>1$), or supplies it with the wrong sign of charge ($\alpha<0$), Ampere's law guarantees the inductive growth of $E_{||}$ which can trigger the runaway production of $e^{\pm}$ 
\citep{beloborodov_polar-cap_2008}. The time-dependent gaps
observed in the simulation differ from
the gap models based on electrostatic considerations (Gauss' law), including the classical polar-gap \citep{ruderman_theory_1975}, slot-gap \citep{arons_pair_1983,muslimov_high-altitude_2004}, and outer-gap \citep{cheng_energetic_1986}.

A time-dependent gap
with $\alpha>1$ forms
near the magnetic axis
above the pulsar polar cap.
This gap occurs because the general-relativistic frame-dragging effect reduces the apparent rotation of the star, and thus reduces $\rho_\text{GJ}$ near the surface, while leaving the required current $j_B$ unchanged \citep{muslimov_tsygan,philippov_ab-initio_2015,belyaev_spatial_2016,Gralla_2016}.
The resulting  $\alpha>1$ renders
the electron flow extracted from the star unable to supply $j_B$, even if it moves at the speed of light.
Thus, $E_{||}$ is induced,
igniting $e^{\pm}$ discharge. The discharge begins when electrons are lifted from the atmosphere by $E_{||}$ and accelerated to energies $\sim \gamma_\text{thr}mc^2$. The electrons emit gamma-rays, which convert to $e^{\pm}$ through the $\gamma-B$ channel.  The increasing density of pairs quickly screens $E_{||}$, and the $e^\pm$ shower proceeds outward. As the cloud of $e^{\pm}$ leaves, $E_{||}$ grows again and the discharge cycle repeats. The discharge reaches maximum multiplicity $\mathcal{M}\equiv n/n_\text{GJ} \sim 10$.
The gap height is 
$h\sim 0.4 r_\star$,
and the cycle time is 
$\sim h/c$.

The magnetosphere prevents the accumulation of net electric charge on the neutron star by arranging positive return-currents, which compensate the negative charge flowing out of the polar-cap (Appendix~\ref{global_currents}). Most of the return-current flows along the thin separatrix layer and is sustained by pair creation at the Y-point \citep{hu_axisymmetric_2021}. We also observe a bulk return-current of macroscopic thickness on open field lines outside the separatrix layer which pass through the null-surface defined by $\rho_\text{GJ} = 0$ [Fig.~\ref{waves} top panel]. The bulk return-current draws electrons from the 
pair plasma created around the light-cylinder.

The simulation reveals two time-dependent gaps on the field lines that
conduct the bulk return-current: (i) an inner return-current gap of type $\alpha<0$ forms above the polar-cap where $\rho_\text{GJ}<0$ and $j_{B}>0$, and (ii) an outer return-current gap of type $\alpha>1$ forms outside the null-surface where $\rho_\text{GJ}>0$ and $j_{B}>0$. The presence of two gaps on the same field lines results in coupled discharge dynamics.
The two interacting discharges reach $e^\pm$ multiplicities ${\cal M}\sim 5$ and repeat on the timescale $\sim 2R_{\rm LC}/c$.

The inner return-current gap has $\rho_\text{GJ}<0$, so here the charge density can only be supplied by electrons. Electrons flow toward the star in order to conduct $j_\text{B}>0$. When this flow dwindles, a gap opens inside the null-surface
(with $E_{||}$ directed away from the star)
and expands 
inward
until it reaches altitude 
$r\sim 2r_\star$
where pair production through the $\gamma-B$ channel 
becomes
possible. 
Then 
discharge 
develops, 
seeded by 
the inward electron flow accelerated in the gap. Pair creation quickly screens $E_\parallel$,
and the shower proceeds toward the star, spreading laterally due to the 
curvature of the magnetic field lines and the finite free paths of tangentially emitted gamma-rays.
Some of the secondary positrons are reversed by $E_\parallel$ at the discharge onset; they escape outward and serve as seeds for the discharge in the outer return-current gap. 

The outer return-current gap extends from the null-surface toward $R_\text{LC}$ (Fig.~\ref{waves}, top). The gap occurs because the required current is positive, $j_{B}>0$, but positrons are not readily available from inside the null-surface where $\rho_\text{GJ}<0$. This generates $E_\parallel$ 
directed away from the star. It pulls electrons into the gap from 
larger radii.
The inward flowing 
electrons supply the correct sign of the current, but the wrong sign of charge ($\rho_\text{GJ}>0$), so they cannot screen the gap. Instead, they get accelerated by $E_\parallel$, cross the null-surface and trigger the inner return-current discharge, which provides a 
source of positrons from inside the null-surface. 
These seed positrons get accelerated and
trigger pair production in the outer return-current gap through the $\gamma-\gamma$ channel. 
Thus, the two gaps (inner and outer) assist each other in 
repeating
pair discharge.

\section{Discharge Waves} 
Fig.~\ref{waves} shows the plasma waves excited during the polar-cap discharge [inset (i)], the inner return-current discharge [inset (ii)], and the outer return-current discharge [inset (iii)]. The electromagnetic fields of the waves $\delta \vec{B}$ and $\delta \vec{E}$ are isolated by first subtracting the time average to remove the zero frequency (background) component. Then we subtract a local spatial average calculated along the magnetic field line, which removes large amplitude low frequency oscillations caused by breathing of the global magnetosphere. The perpendicular part of the waves are calculated as $\delta \vec{B}_{\perp} = \vec{B} - \delta \vec{B}_{||}$, where $\delta \vec{B}_{||}=\delta \vec{B}\cdot\vec{B}/|\vec{B}|$.

The waves are remarkably consistent with a mechanism previously seen in  local-box simulations \citep{philippov_origin_2020,cruz_coherent_2021}: pair production in the gap drives a variable current which couples to oscillations of $\delta{\vec{E}}$ and $\delta \vec{B}$.
Two properties of the observed $\delta\vec{E}$ show that it is the superluminal O-mode \citep{1986ApJ...302..120A}:  $\delta\vec{E}$ lies in the $\vec{k}$-$\vec{B}$ plane ($\vec{k}$ is the wavevector) [Fig.~\ref{waves}], and $\delta{E}_\parallel/\delta E_\perp$ is consistent with the superluminal O-mode polarization (Appendix~\ref{normal_modes}).
Excitation occurs with  frequencies $\omega$ comparable to the plasma frequency $\omega_p=(4\pi n e^2/m)^{1/2}$, which is rapidly growing during the discharge \citep{tolman_electric_2022}.
The growing $\omega\sim\omega_p$ and  the moderate $k$ (large length-scale of the discharge) lead to the 
excitation of the superluminal branch $\omega/k>c$ rather than subluminal Alfv\'en waves.
The waves reach amplitude $E^{\star}_{||}\sim \gamma_s m c\,\omega_\star / e $, where $\omega_\star$ is the plasma frequency later in the discharge when screening reduces $E_{||}$ so that it
can barely reverse
$e^{\pm}$ \citep{tolman_electric_2022}.
The emitted waves track adiabatically along the O-mode branch as they propagate through the decreasing plasma density, and should escape the magnetosphere as vacuum waves with $k_\star=\omega_\star /c$ (the escape is not captured in our simulation, because at large radii the grid cell  exceeds $k_\star^{-1}$.)
The wave emission is strongly modulated on the gap light-crossing timescale.

Our simulation also demonstrates the extraordinary (X) modes emitted from 
the equatorial plasmoid chain [Fig.~\ref{waves}, insets (iii), (iv)] \citep{philippov_pulsar_2019,lyubarsky_2019}. These waves have $\delta\vec{E}\parallel (\vec{k}\times\vec{B})$ and show up only in $\delta\vec{E}_\perp$ with no contribution to $\delta E_\parallel$. Note
that in axisymmetry $\vec{k}$ lies in the poloidal plane. Near the star the background magnetic field $\vec{B}$ is approximately poloidal.
At large radii $\vec{B}$ is twisted out of the poloidal plane.

\section{Plasma Instabilities} 

\begin{figure}[h!]
\centering
\includegraphics[width=.47\textwidth]{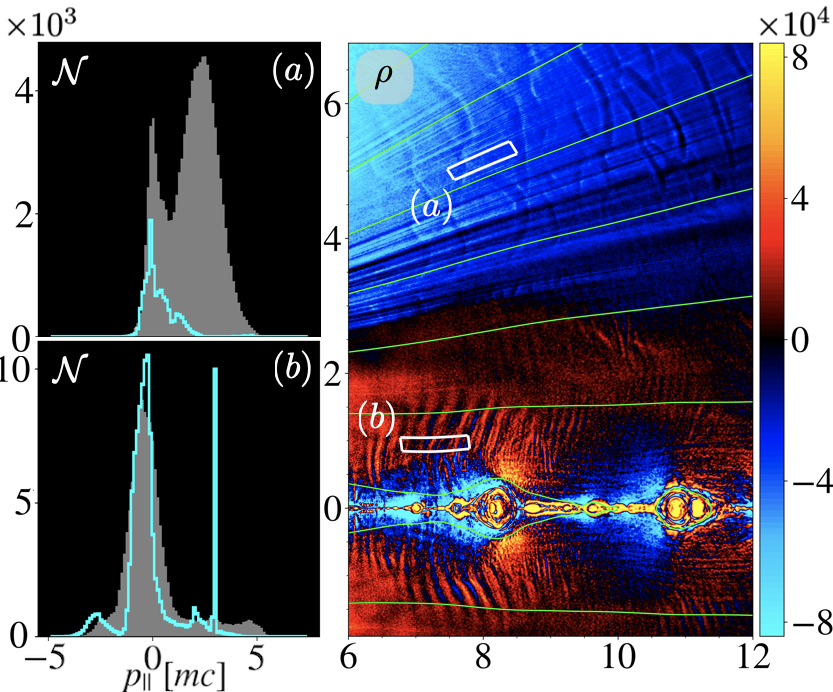} 
\caption{Beam instabilities in the outer magnetosphere. Left: Parallel momentum ($p_{||}$) distributions of $e^{\pm}$ (grey) and gamma-rays (cyan) in regions (a) and (b) in the right panel. For gamma-rays we show distributions of $p_{||}/2$ (their momentum is split between the created $e^{+}$ and $e^{-}$).
(a): Gamma-rays from the current-sheet (cyan peak) inject $e^{\pm}$ (grey peak). (b): Gamma-rays from the separatrix (thin cyan peak) and current-sheet (wider cyan peak) inject different $e^{\pm}$ beams. Right: Charge density $\rho$ and charge bunches.}
\label{beam}
\end{figure}

The simulation also reveals 
beam plasma
instabilities, triggered at several sites in the outer magnetosphere. The triggering mechanism is non-local: the $e^\pm$ wind becomes exposed to a gamma-ray beam emitted elsewhere in the magnetosphere, which injects an $e^\pm$ beam with a different momentum (Fig.~\ref{beam}). Both streams are made of
secondary 
$e^\pm$,
and thus the instability growth rate is not strongly suppressed by 
ultra-high Lorentz factors typical for 
primary accelerated particles.
The instability is electrostatic in nature, and controlled by plasma oscillations along the magnetic field lines (rapid gyration of the particles increases their effective parallel inertia, but does not change the qualitative behaviour of the instability). Charge bunches form at each site where the instability 
occurs
(Fig.~\ref{beam}).

The beam energy is deposited into growing plasma modes 
seen in Fig.~\ref{waves}, inset (iv).
The modes grow in the outflowing wind over a characteristic distance $d$ consistent
with the growth rate of the relativistic beam instability $d \sim c/\Gamma$
(Appendix~\ref{beam_growth}) \citep{egorenkov_beam_1983}. The measured wavelength of the modes is consistent with $\lambda_0\sim 2\pi c / (\omega_p \gamma^{1/2})$, indicating that the instability is first triggered on Cherenkov resonance \citep{egorenkov_beam_1983}.
The modes are inclined to the local magnetic field, and have transverse and longitudinal components [Fig.~\ref{waves}, inset (iv)]. The phase speed is necessarily subluminal. Therefore, we identify the excited waves as hybrid electrostatic-Alfv\'en modes
\citep{1986ApJ...302..120A,Rafat_2019}. The waves have amplitude $E_{||}\sim 4\pi \Delta\rho \lambda_0$, where $\Delta\rho \sim \rho_\text{GJ}$ is the charge density of the observed bunches (Fig.~\ref{beam}). 
The fate of the waves and 
radiation 
from
the charge bunches should be investigated with detailed local simulations.

\section{Discussion} 
Our numerical experiment has achieved
a voltage 
high
enough to reveal 
the locations where the $e^{\pm}$ discharge 
ignites and produces
coherent radio waves. In addition to the well known 
polar-cap gap, the simulation
has uncovered a pair of interacting gaps
in the outer magnetosphere,
which form around the null-surface in the bulk return-current.
Furthermore, we observed
waves and charge bunches excited by beam instabilities
in the secondary $e^\pm$ plasma --- a result of multiple gamma-ray streams revealed by the global simulation. The high resolution is key to these findings, providing a glimpse of the origin of pulsar radio waves, a subtle phenomenon compared to pulsar gamma-ray emission.

The electric discharge in our simulation reaches pair multiplicity $\mathcal{M}\sim 5-10$,
much smaller than in
real pulsars. However,
it is large enough to demonstrate the 
limit-cycle behaviour of the gaps and the excitation of electromagnetic modes in the global magnetosphere. 
Realistic multiplicities up to $\mathcal{M}\sim 10^4$ are so far achieved only in local 1D discharge simulations \citep{timokhin_2010, timokhin_current_2013},
and
excitation of waves 
in a polar-cap discharge
was previously demonstrated in local 2D simulations 
with ${\cal M}\sim 10$
\cite{philippov_origin_2020,cruz_coherent_2021}. 
Simple estimates suggest that the discharge waves 
offer a promising mechanism for pulsar radio emission. The waves are excited near the local plasma frequency $\nu\sim\nu_p$ which scales as $\mathcal{M}^{1/2}$. For a realistic $\mathcal{M}$, one can estimate $\nu \sim (2\pi)^{-1}(4\pi \mathcal{M} n_\text{GJ} e^2 / \langle \gamma^3 \rangle m)^{1/2}\sim$~MHz-GHz \citep{timokhin_polar_2015,philippov_origin_2020}. The expected wave amplitude at the polar-cap $E^\star_{||}\sim 10^5~(\gamma_s/10^3)(\nu_\star /\text{GHz})$~G implies $L_\text{radio}\sim c {E^{\star}_{||}}^2 A_\text{pc} / (4\pi) \sim 10^{29}~(A_\text{pc}/10^9\text{cm}^2)~\text{erg~s}^{-1}$ \citep{tolman_electric_2022}, similar to observed pulsars ($A_\text{pc}$ is the polar-cap area).

The global kinetic simulation offers a physical framework to start interpreting the rich pulsar observations.
In particular, the polar-cap discharge may be responsible for the so-called ``core'' radio emission, and 
the inner return-current discharge may produce ``conal" radio emission. 
Their strong modulation by the electric discharge of the gaps 
may produce microstructure in the pulse profiles. Also interesting are the additional plasma instabilities found in the outer magnetosphere, which are special to energetic pulsars with active pair creation around the light cylinder.

\balance
However, significant work is needed to improve the simulations before they can be directly compared to individual pulsars.   
In particular, increasing multiplicity $\mathcal{M}$ may change the interaction of the two return-current gaps: instead of assisting each other with seed particles for discharge, the inner gap with a high $\mathcal{M}$ may flood the outer gap and switch off its radio emission.
A more detailed implementation of
gamma-ray emission and $e^{\pm}$ production 
may
change the 
beam instabilities and charge bunching. 
It has been suggested that 
charge bunches could produce coherent curvature emission in the radio band \citep{ruderman_theory_1975,Gil_2004} 
and the waves experience propagation effects \citep{barnard_1986,Lyubarsky_1993,Lyubarskii_1998, beskin_2012}, which needs further investigation.
Finally, our simulation was 
limited to aligned rotators $\boldsymbol{\mu}\parallel\boldsymbol{\Omega}$. It will be 
essential 
to investigate 
radio emission in 
inclined rotators. 
\\

\section*{acknowledgments}
The authors thank A. Chen, R. Hui, A. Philippov, and M. Ruderman for useful discussions. Resources supporting this work were provided by the NASA High-End Computing Program through the NASA Advanced Supercomputing Division at Ames Research Center. A.B. is supported by a PCTS fellowship and a Lyman Spitzer Jr. fellowship.
A.M.B. acknowledges grant support by NSF AST 2009453, NASA 21-ATP21-0056, and Simons Foundation \#446228.
Y. L. is supported by NSF Grant AST-2009453.

\appendix

\onecolumngrid

\section{Global Currents and Pair Production}
\label{global_currents}

\begin{figure*}[h!]
\centering 
\includegraphics[width=\textwidth]{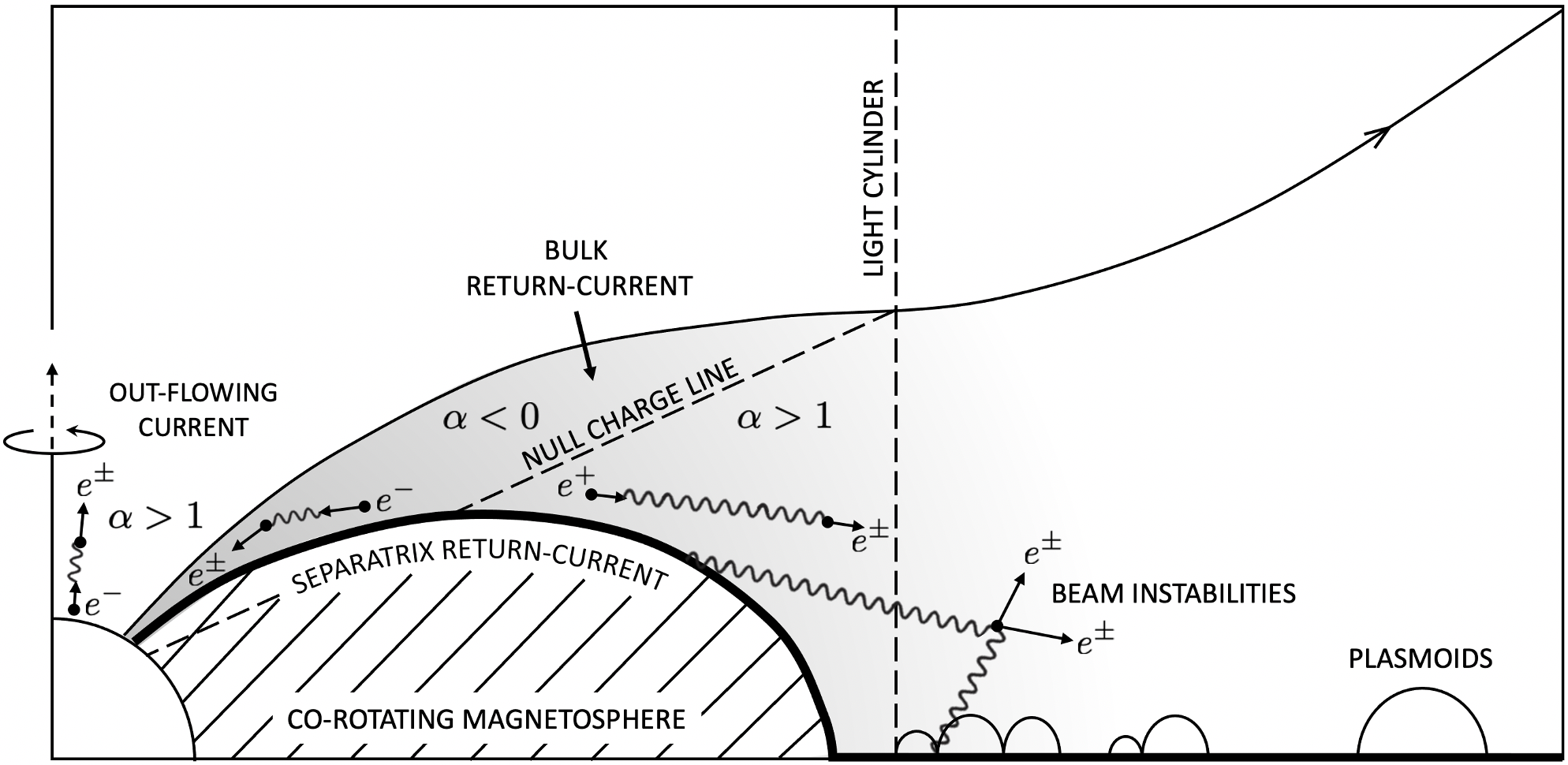} 
\caption{Global currents and $e^\pm$ production in the aligned rotator. The white region above the polar cap is
the zone of out-flowing current, the shaded grey region indicates the bulk return-current, and the thick black curve shows
the separatrix return-current layer. The hatched area indicates the closed zone. The null charge line (dashed) is defined by $\rho_\text{GJ}=-\vec{\Omega}\cdot \vec{B}/(2\pi c)=0$; above the line $\rho_\text{GJ}<0$, and below $\rho_\text{GJ}>0$. Episodic $e^\pm$ discharge occurs in the out-flowing current above the polar cap ($\alpha>1$), in the bulk return-current above the polar cap ($\alpha<0$), and in the bulk return-current outside the null line ($\alpha>1$). Beam instabilities occur outside the light cylinder where $e^\pm$ are created by the intersecting gamma-ray beams with different directions.
}
\label{diagram}
\end{figure*}

Figure \ref{diagram} summarizes the magnetospheric structure and the discharges that fill it with plasma, as observed in our simulation.
The location of the discharges is controlled by the global magnetospheric currents, and occurs where $\alpha \equiv j_B/(c\rho_\text{GJ})>1$ or $\alpha<0$. 
The discharge in the out-flowing current above the polar cap (Fig.~\ref{diagram}, white region) is well known; it  develops because here $\alpha$ is increased above unity by the frame dragging effect in the spacetime of the rotating star \citep{muslimov_tsygan}. Discharge in the return-current zone is more subtle. Approximately 90\% of the return current flows in the separatrix current sheet, and 10\% flows in the grey region (Fig.~\ref{diagram}), filling a significant volume with  $\alpha<0$ near the star. In this ``bulk'' return-current region above the polar cap, we observe strong episodic $e^\pm$ discharges. Similar discharges with $\alpha<0$ were previously simulated in one-dimensional (plane-parallel) models \citep{timokhin_current_2013}. The discharge observed in our simulation, however, operates in the `thin-tube' regime (the transverse size of the electric gap is smaller than its length along $\vec{B}$) and does not admit a plane-parallel description. The discharge is seeded by the time-dependent plasma supply from larger radii, and the particles accelerate toward the star  with secondary pair production, in repeating episodes.
The separatrix return-current (Fig.~\ref{diagram}, thick black curve)
has $\alpha<0$
 and is 
sustained by 
pair
production at the Y-point
as explained in \citep{hu_axisymmetric_2021}.
Pair production 
along and within the separatrix itself
is inefficient because 
it is very thin and curved, so the emitted gamma-rays
quickly travel outside its thickness before converting to $e^{\pm}$.
As a result,
the separatrix return-current does not display the cyclic $e^\pm$ discharge episodes.

\section{Dispersion Relation and Normal Modes of Pulsar Plasma}
\label{normal_modes}

Here we review the dispersion relation and normal modes of the $e^{\pm}$ pair plasma expected to exist in pulsar magnetospheres. Near the neutron star surface the plasma is effectively infinitely magnetized. Therefore, charged particles only move parallel to the magnetic field, and the plasma supports a very limited set of normal modes \citep{1986ApJ...302..120A}. The dispersion relation is obtained from the Vlasov-Maxwell equations \citep{1986ApJ...302..120A},
\begin{equation}
    (\omega^2 - c^2 k^2)\left[ (\omega^2 - c^2 k_{||}^2)\left( 1 - \frac{\omega_p^2}{\omega^2} g \right) - c^2 k_\perp^2 \right]= 0,
    \label{dispersion}
\end{equation}
where $\omega$ is the wave frequency, $\omega_p$ the plasma frequency, $k_{||}$ and $k_\perp$ are the components of the 
 wavevector parallel and perpendicular to $\vec{B}$, $g=\langle \gamma^{-3} \left(1 - kv/\omega\right)^{-2}\rangle$, and $\langle ...\rangle$ signifies
an average over particle momenta. If the plasma is cold, $g=1$ in the plasma rest frame. The dispersion relation describes three modes: The extraordinary (X) mode with $\omega=ck$ and electric field polarized in the $\vec{k}\times\vec{B}$ direction, and two branches of the ordinary (O) mode with electric field polarized in the $\vec{k}-\vec{B}$ plane. In the dense discharge plasma ($ck/\omega_p \ll 1$), 
the polarization of the two branches of the O-mode is described by
\begin{equation}
    \frac{\delta E_{||}}{\delta E_\perp} \approx 
    \begin{cases}
    \begin{aligned}
      &-\frac{\omega_p^2}{c^2k_{||}k_\perp} \text{\quad\quad superluminal O-mode, }\\ 
      &~~~~\frac{c^2k_{||}k_\perp}{\omega_p^2} \text{\quad\quad subluminal O-mode (Alfv\'en mode)}, 
    \end{aligned}
  \end{cases}
\end{equation}
where $\delta E_{||}$ is the wave electric field parallel to $\vec{B}$, and $\delta E_\perp$ is the wave electric field perpendicular to $\vec{B}$ in the $\vec{k}-\vec{B}$ plane \citep{1986ApJ...302..120A}. The Alfv\'en 
branch
has phase speed $\omega/k<c$, the superluminal O-mode mode $\omega/k>c$, and the extraordinary mode has  $\omega/k=c$. For a given $\vec{k}$, the three modes (extraordinary, Alfv\'en, and superluminal) are orthogonal. The Alfv\'en mode suffers from Landau damping, and cannot escape the magnetosphere without non-linear conversion to a different eigenmode
\citep{1986ApJ...302..120A}. 

By contrast, as the superluminal mode propagates outward through the decreasing plasma density, it tracks adiabatically along the superluminal branch and becomes a freely propagating electromagnetic wave \citep{1986ApJ...302..120A,barnard_1986}. 
During the propagation the wave frequency $\omega_\star$ is fixed, while $k$ increases and approaches $k_\star =\omega_\star /c$ (the phase speed approaches $c$ from above). 
The superluminal mode is attractive for pulsar radio emission because it does not suffer from Landau damping ($\omega/k >c$), and it can escape the magnetosphere as a vacuum electromagnetic wave in the radio band \citep{philippov_origin_2020}.

\section{Beam Instability in Pulsar Plasma}
\label{beam_growth}

Here we review the relativistic beam instability in highly magnetized $e^{\pm}$ plasma (for a detailed calculation see \citep{egorenkov_beam_1983}). Consider a uniform background of $e^{\pm}$ with uniform density $n_p$ which streams relativistically along the magnetic field with Lorentz factor $\gamma$. The beam has uniform density $n_b<n_p$, and a Gaussian distribution of momentum parallel to the magnetic field with mean $p_b=\gamma_b mc$, and width $\Delta p_{b} = \Delta \gamma_{b} mc $. If the beam density is sufficiently high 
or $\Delta \gamma_{b}$ is sufficiently small, the instability operates in the hydrodynamic regime with a growth rate $\Gamma\gg kc\,\Delta \gamma_b/\gamma_b^3$. 
Then, the growth rate can be found by setting $\Delta \gamma_{b}\rightarrow 0$. In particular, for purely longitudinal oscillations 
the dispersion relation 
has the form (Eq.~\ref{dispersion})
\begin{equation}
    1 - \frac{\omega_p^2}{\omega^2}g - \frac{\omega_b^2}{\omega^2}\frac{1}{ \gamma_b^3 (1 - kv_b/\omega)^2} = 0,
\end{equation}
where the third term is due to the beam with $\omega_b=\sqrt{4\pi n_b e^2 / m}$ and $v_b = c
(1 + \gamma_b^{-2})^{-1/2}$ 
\citep{egorenkov_beam_1983}. Near Cherenkov resonance, the dispersion relation can be written in the form $\omega = kv_b + \Delta \omega$, with $|\Delta \omega|\ll k v_b$. The growth rate is maximal when $k$ approaches $ k_0 \equiv 2\omega_p \langle \gamma \rangle ^{1/2}/c$, 
and is given by
\begin{equation}
    \Gamma = \text{Im}~\Delta \omega =\frac{\sqrt{3}}{4 \cdot 2^{1/3}}\left( \frac{n_b}{n_p}\right)^{1/3}\frac{ck_0}{\gamma_b\langle \gamma^3 \rangle^{1/3}}.
    \label{rate}
\end{equation}
The condition for the instability to operate in the hydrodynamic regime is then
\begin{equation}
    \left( \frac{n_b}{n_p}\right)^{1/3}\gg \frac{\langle \gamma\rangle  \Delta\gamma_b}{\gamma_b^2}.
    \label{hydro}
\end{equation}
This condition may be satisisfied even when the momentum spread of the beam is large, $\Delta \gamma_b \gtrsim \gamma_b$.
Note that the instability also occurs at $k\ll k_0$, although 
at a slower rate \citep{egorenkov_beam_1983}. 
The beam instability grows only the subluminal (Alfv\'en)
mode; the superluminal mode cannot satisfy the Cherenkov resonance $\omega/k=v_b$.

Beam instabilities in pulsar magnetospheres are usually discussed in the context of primary accelerated $e^{\pm}$ with $\gamma_b\sim 10^6$ (beam) interacting with the secondary $e^{\pm}$ with $\gamma \sim 10^2$-$10^3$ (background plasma) \citep{ruderman_theory_1975}. The growth rate (Eq.~\ref{rate}) is then strongly suppressed by the high $\gamma_b$, resulting in $\Gamma\ll\Omega$, where $\Omega$ is the rotation rate of the pulsar. The plasma escapes the magnetosphere on the timescale $r_{\rm LC}/c=\Omega^{-1}$, and so there is not enough time for the instability to develop.

By contrast, our simulation reveals locations where both the background plasma and the beam consist of secondary $e^\pm$: there is a wind of $e^\pm$ produced near the star and flowing out to the light cylinder and $e^\pm$ beams injected by the gamma-rays from the equatorial current sheet. The moderate Lorentz factors of the created  $e^\pm$,  $\gamma_b\sim 10^2$-$10^3$, imply a fast growth of the beam instability $\Gamma>\Omega$. Indeed, using a typical $\gamma\sim 10^2$ for the wind and $\gamma_b\sim 10^3$ for the injected beam, we find
\begin{equation}
    \frac{\Gamma}{\Omega} \sim \left(\frac{n_b}{n_p}\right)^{1/3}\frac{\omega_p}{\Omega \gamma_b \gamma^{1/2}} \sim 10^2 ~ \left( \frac{\gamma_b}{10^3} \right)^{-1} \left( \frac{\gamma}{10^2} \right)^{-1/2} \left(\frac{P}{0.033~\text{s}} \right)^{1/2} \left( \frac{B_\text{LC}}{10^{6}~\text{G}}\right)^{1/2} \left( \frac{\mathcal{M}}{10^{2}}\right)^{1/2},
\end{equation}
where ${\cal M}=n_p/n_{\rm GJ}$ is the pair multiplicity, $P=2\pi/\Omega$, $B_\text{LC}$ is the magnetic field strength at the light-cylinder, and we set $n_b/n_p\sim 0.1$. We conservatively normalized ${\cal M}$ to a modest value of $10^2$, and a higher ${\cal M}$ will only increase $\Gamma/\Omega$. The beam instabilities require pair production by photon-photon collisions around the equatorial current sheet, which occurs in rapidly rotating pulsars; therefore, we normalized $B_{\rm LC}$ to $10^6$\,G, as found e.g. in the Crab pulsar.
Note also that for extremely energetic pulsars the $e^\pm$ created around the current sheet may dominate the wind density and serve as the main plasma background; then, the wind emerging from the inner magnetosphere will play the role of the beam.

The beam instabilities around the equatorial current sheet will lead to a new component of pulsar radio emission if there is significant conversion of the generated subluminal O-mode to another eigenmode, which can escape.

\bibliographystyle{apj}
\bibliography{Astrophysics}

\end{document}